\documentclass[%
reprint,
aps,
pra,
letterpaper,
]{revtex4-2}

\usepackage[utf8]{inputenc}
\usepackage{amsmath,amsfonts,amssymb}
\usepackage{mathtools}
\usepackage[colorlinks,unicode]{hyperref}
\usepackage{enumitem}
\usepackage{dsfont}

\newcommand{\eye}{\operatorname{i}}
\newcommand{\op}[1]{\ensuremath{#1}}
\newcommand{\map}[1]{\ensuremath{\mathbf{#1}}}
\newcommand{\textgate}[1]{\text{\texttt{#1}}}
\newcommand{\vac}[1]{\ensuremath{\tilde{#1}}}
\newcommand{\e}{\operatorname{e}}
\newcommand{\aket}[1]{\big| #1 \big>} 
\newcommand{\abra}[1]{\big< #1 \big|} 
\newcommand{\abraket}[2]{\big< #1 \vphantom{#2} \big| \big. #2 \vphantom{#1} \big>} 
\newcommand{\Time}{\ensuremath{t}}
\newcommand{\TimeGeneral}{\ensuremath{t}}
\newcommand{\DeltaTime}{\ensuremath{\Delta t}}
\newcommand{\OrthogonalisationTime}{\ensuremath{t_\perp}}
\newcommand{\TimeInbound}{\ensuremath{t_\mathrm{in}}}
\newcommand{\TimeOutbound}{\ensuremath{t_\mathrm{out}}}
\newcommand{\TimeGeneralFirst}{\ensuremath{t'}}
\newcommand{\TimeGeneralSecond}{\ensuremath{t''}}
\newcommand{\CR}{\mathrm{CR}}
\newcommand{\CV}{\mathrm{CV}}
\newcommand{\CRState}{\ensuremath{\sigma}}
\newcommand{\CVState}{\ensuremath{\theta}}
\newcommand{\EvolutionA}{{\ensuremath{A}}}
\newcommand{\EvolutionB}{{\ensuremath{B}}}

\newcommand{\DCTCsCRMap}{\ensuremath{D}}
\newcommand{\DCTCsCVMap}{\ensuremath{T}}
\newcommand{\PCTCsMap}{\ensuremath{P}}

\newcommand{\NumberLevels}{\ensuremath{N}}
\newcommand{\Normalisation}{\ensuremath{\mathcal{N}}}
\newcommand{\Unitary}{\ensuremath{U}}
\newcommand{\timeunitary}{\ensuremath{R}}
\newcommand{\Identity}{\mathds{1}}
\newcommand{\Rotation}{\ensuremath{R}}
\newcommand{\Swap}{\ensuremath{S}}
\newcommand{\UnitaryVacuum}{\ensuremath{U}}
\newcommand{\UnitaryTraced}{\ensuremath{W}}
\newcommand{\qclock}{\ensuremath{\phi}}
\newcommand{\Free}{\ensuremath{g}}
\newcommand{\Vacuum}{\ensuremath{\Omega}}
\newcommand{\DeltaEnergy}{\ensuremath{\Delta E}}
\newcommand{\MapBracket}[1]{\big[ #1 \big]}
\newcommand{\MapBrackets}[2]{\big[ #1 \vphantom{#2} , #2 \vphantom{#1} \big]}

\newcommand{\abs}[1]{\left| #1 \right|}
\newcommand{\norm}[1]{\left| \left| #1 \right| \right|}
\newcommand {\tr} {\mathrm{tr}}

\begin{document}

\title{Time-travelling billiard-ball clocks: a quantum model}

\author{Lachlan G. Bishop}
\email{lachlan.bishop@uq.net.au}
\author{Fabio Costa}
\author{Timothy C. Ralph}
\affiliation{%
	School of Mathematics and Physics, The University of Queensland, St. Lucia, Queensland 4072, Australia
}%

\date{\today}

\begin{abstract}
	General relativity predicts the existence of closed timelike curves (CTCs), along which an object could travel to its own past. A consequence of CTCs is the failure of determinism, even for classical systems: one initial condition can result in multiple evolutions. Here we introduce a quantum formulation of a classic example, where a billiard ball can travel along two possible trajectories: one unperturbed and one, along a CTC, where it collides with its past self. Our model includes a vacuum state, allowing the ball to be present or absent on each trajectory, and a clock, which provides an operational way to distinguish the trajectories. We apply the two foremost quantum theories of CTCs to our model: Deutsch's model (D-CTCs) and postselected teleportation (P-CTCs). We find that D-CTCs reproduce the classical solution multiplicity in the form of a mixed state, while P-CTCs predict an equal superposition of the two trajectories, supporting a conjecture by Friedman \emph{et al.}~[Phys.~Rev.~D \textbf{42}, 1915 (1990)].
	\newline
\end{abstract}

\maketitle

\section{\label{sec:introduction}Introduction}

Closed timelike curves (CTCs) form an interesting class of objects within the general theory of relativity as they present the possibility for an observer to travel back in time. Consequently, any physical system may be able to interact with its past self, and so the potential existence of CTCs within our own universe naturally evokes scientific investigation into time travel. The foremost focus of such research is on questions regarding the nontrivial causal structure of CTC spacetimes and the consistency of the standard laws of physics.

Time-travel paradoxes and the concept of retrocausality lie at the foundation of the issues with CTCs. One way to explore the compatibility between the laws of physics and these exotic objects is to determine whether the two can be consolidated without having unacceptable causality violations manifest. This is typically accomplished by studying the evolution of simple physical models in spacetimes containing CTCs. In these problems, one necessitates that a system, while allowed to propagate into its own past, must do so in a way that is consistent with its own original history. Any interaction that occurs must be compatible with the past, and causal sequences containing events solely of this nature are characterised as being self-consistent. The \emph{principle of self-consistency} \cite{friedman_cauchy_1990}, an innate law by which the universe is conjectured to operate, serves to suppress temporal paradoxes via the prohibition of pathological causal sequences. Under such a condition, a globally consistent solution of the local physical laws must exist, or else the problem is ill-posed.

\subsection{\label{sec:introduction_billiard}Time-travelling billiard balls}

For a spacetime with one or more chronology-violating sets (i.e., time machines), the past can be influenced by the future, provided that any changes which are made obey the self-consistency principle. The lack of such changes (i.e., self-interactions) in the Cauchy problem for noninteracting classical fields seems like a probable cause of its well-posedness on CTC spacetimes \cite{friedman_cauchy_1990,friedman_cauchy_1991,friedman_existence_1997,bachelot_global_2002,friedman_cauchy_2004,arefeva_cauchy_2008,volovich_solutions_2009,groshev_existence_2010,bachelot_kleingordon_2011,bullock_klein-gordon_2012}. Alternatively, a system that is able to self-interact (e.g., collide with its past or future self) is more likely to display unusual results\textemdash indeed, with an interacting field, uniqueness is thought to be lost as the strength of the interaction increases \cite{friedman_cauchy_2004}. Therefore, in order to study this, models of interacting systems near CTCs were developed, and perhaps the most famous of these considers the elastic self-collision(s) of a billiard ball that exists in a wormhole-based time machine spacetime \cite{morris_wormholes_1988,novikov_analysis_1989,thorne_laws_1991,lossev_jinn_1992,ralph_relativistic_2012}. In this framework, a solid, elastic, spherical mass (the ``billiard ball'') enters one mouth (the future mouth) of a wormhole, exits the other (the past mouth) at an earlier time (due to a time shift having been induced between the two), and then collides with its earlier (past) self. Depending on the initial position and velocity of the ball, drastically distinct evolutions of the ball through the CTC-wormhole region can arise.

\begin{figure}[b]
	\includegraphics{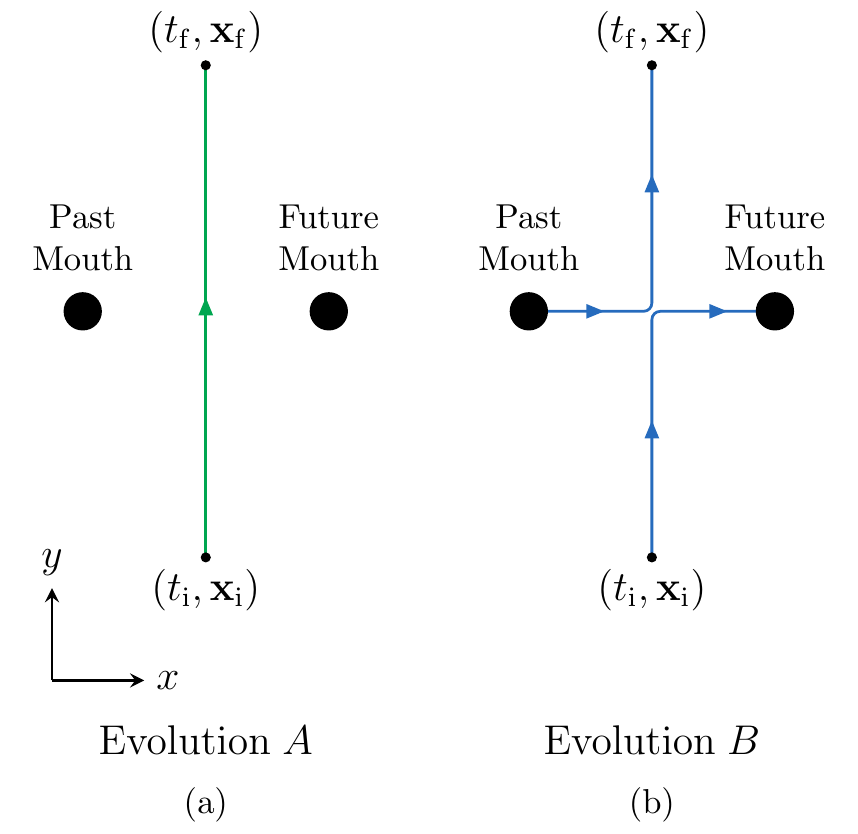}
	\caption[Classical billiard ball problem]{\label{fig:billiard_balls}Two possible spatial trajectories of a classical billiard ball through CTC-wormhole spacetimes with the same initial data (i.e., position and velocity) in a two-dimensional slice of three-dimensional space. (a) A chronology-respecting history (evolution $\EvolutionA$) in which the billiard ball evolves freely between the CTC and wormhole; (b) a chronology-violating history (evolution $\EvolutionB$) in which the billiard ball is struck onto the CTC path by its future self, thereby causing it to time travel into its past and subsequently strike its past self onto the CTC. The initial and final conditions for both evolutions are, respectively, exactly the same, meaning that the histories are (classically) indistinguishable.}
\end{figure}

Studies \cite{friedman_cauchy_1990,echeverria_billiard_1991,lossev_jinn_1992,novikov_time_1992,mikheeva_inelastic_1993,mensky_three-dimensional_1996,dolansky_billiard_2010} involving time-travelling billiard balls have shown that given the same initial data posed in the presence of CTCs, there can be multiple self-consistent solutions which satisfy the equations of motion. Figure \ref{fig:billiard_balls} illustrates a prominent example in which there are (at least) two distinct histories through which the billiard ball may evolve in a CTC-wormhole spacetime. This scenario is the foremost example of the \emph{(simplified) billiard-ball paradox} (due originally to Thorne in \cite{friedman_cauchy_1990}), and will hereafter be referred to as such. Unusual for time-travel paradoxes, the paradoxical issue here is not of self-inconsistent trajectories, but is of indeterminism in the self-consistent ones. This is to say that solutions always exist because they self-adjust themselves (thereby providing consistency), but one subsequently faces a different issue: solution multiplicity. This new problem is interesting, as the existence of more than one solution to the equations of motion contrasts with the determinism typically associated with classical mechanics.

It is natural to question whether the characteristic of solution multiplicity is indeed merely an artefact of a classical description and vanishes in a fully self-consistent quantum mechanical treatment, or whether it remains as a pathological aspect of CTCs and time travel under any physical description. In past work \cite{friedman_cauchy_1990}, it has been suggested that the indeterminism problem of the billiard-ball paradox disappears in quantum theory. The reasoning behind this is that unlike classical mechanics, a quantum treatment provides one with ways of assigning solution probabilities, thereby replacing the classical theory's multiplicity of solutions with a set of probabilities for the outcomes of all sets of measurements.

For example, a quantum prescription based on the path-integral or sum-over-histories formulation tells us that classical trajectories are probabilistically superposed. In a prominent paper by Friedman \emph{et al.}~\cite{friedman_cauchy_1990}, the authors propose to apply such a prescription to an idealised scenario where there are exactly two solutions corresponding to one initial condition. When the ball begins in a nearly classical wave packet, Friedman \emph{et al.}~postulate that a Wentzel-Kramers-Brillouin (WKB) approximation to their sum-over-histories method would find that the wave packet emerges from the CTC-wormhole region having travelled along either evolution $\EvolutionA$ or $\EvolutionB$ with equal (i.e., $\frac{1}{2}$) probability. One interpretation of this is to say that the associated quantum billiard ball travels along \emph{both} paths simultaneously, thus ameliorating the indeterminism of classical mechanics.

It is important to note, however, that Friedman \emph{et al.}~do not specify how to extract probabilities from the scenario considered: the two trajectories coincide not only in their initial state, but also in their final state. Therefore, a measurement of the ball's final position would reveal the same outcome, regardless of the path taken. Indeed, no calculations are provided in Ref.~\cite{friedman_cauchy_1990} or subsequent papers to support their WKB approximation conjecture. Nonetheless, the pioneering method proposed by Friedman \emph{et al.}~motivates further study of this problem, as the interesting probabilistic result they postulated is simply inaccessible to classical mechanics.

\subsection{\label{sec:background}Quantum mechanical models of time travel}

Naturally, the advent of CTCs in the semiclassical general theory of relativity prompted study into quantum theories of time travel. In the absence of a complete quantum theory of gravity, reconciling CTCs with standard quantum mechanics forms a compelling basis for research. Indeed, exploration into the interplay between quantum mechanics and CTCs, even in only a theoretical manner, may provide insight into a yet unknown full theory of quantum gravity. In any case, research into this area has so far taken two main routes. The first, in which the principle of self-consistency is applied to the density matrix itself, gives the \emph{Deutsch model} (D-CTCs) \cite{deutsch_quantum_1991}. The second, which is equivalent to a path-integral formulation \cite{politzer_path_1994}, gives \emph{postselected teleportation} (P-CTCs) \cite{lloyd_quantum_2011,lloyd_closed_2011}. Of course, alternative prescriptions of quantum time travel do exist \cite{greenberger_quantum_2005, allen_treating_2014, araujo_quantum_2017, czachor_time_2019, baumeler_reversible_2019, tobar_reversible_2020}, but none are as well-developed as either D-CTCs or P-CTCs and so they will not be discussed in this paper.

For D-CTCs, the mathematical formulation of self-consistent solution(s) is provided by requiring that any given chronology-violating (CV) system enters the wormhole in the state $\op{\CVState}\in\mathcal{H}_\CV$, and then emerges in the past in the same state despite having interacted with a chronology-respecting (CR) system in the (pure) state,
\begin{equation}
	\op{\CRState} = \aket{\psi}\abra{\psi} \in \mathcal{H}_\CR \label{eq:Deutsch_pure}
\end{equation}
through a unitary $\op{\Unitary}\in\mathcal{H}_\CR\otimes\mathcal{H}_\CV$. Self-consistent solution(s) to the evolution of the CTC state may then be identified as fixed point(s) of the CV map, which are expressible via
\begin{equation}
	\op{\CVState} = \map{\DCTCsCVMap}_{\Unitary}\MapBrackets{\op{\CRState}}{\op{\CVState}} = \tr_\CR\left[\op{\Unitary}\left(\op{\CRState} \otimes \op{\CVState}\right)\op{\Unitary}^\dagger\right] \label{eq:Deutsch_CV_2}.
\end{equation}
This condition essentially codifies the requirement that the ``younger'' CTC state (the right-hand side) exiting the gate is the same as the `older' CTC state (the left-hand side) entering the gate. Once solutions to Eq.~(\ref{eq:Deutsch_CV_2}) are determined, the evolution of the system state $\op{\CRState}$ through the interaction $\op{\Unitary}$ in this Deutsch model is then simply given by
\begin{equation}
	\map{\DCTCsCRMap}_{\Unitary}\MapBrackets{\op{\CRState}}{\op{\CVState}} \equiv \tr_\CV\left[\op{\Unitary}\left(\op{\CRState} \otimes \op{\CVState}\right)\op{\Unitary}^\dagger\right]. \label{eq:Deutsch_CR}
\end{equation}

For P-CTCs, on the other hand, the evolution $\op{\CRState}\rightarrow\map{\PCTCsMap}_{U}\MapBracket{\op{\CRState}}$ of the system input state is given by
\begin{equation}
	\map{\PCTCsMap}_{\Unitary}\MapBracket{\op{\CRState}} = \frac{\op{\UnitaryTraced} \op{\CRState} \op{\UnitaryTraced}^\dagger}{\tr\left[\op{\UnitaryTraced} \op{\CRState} \op{\UnitaryTraced}^\dagger\right]} \sim \frac{\op{\UnitaryTraced} \aket{\psi}}{\sqrt{\norm{\op{\UnitaryTraced} \aket{\psi}}}}, \label{eq:P-CTCs_map}
\end{equation}
where the relation operator $\sim$ indicates the pure state form of the preceding density expression and $\op{\UnitaryTraced} \equiv \tr_\CV\left[\op{\Unitary}\right]$ is the partial trace over the Hilbert space of the system in the CTC. Note that P-CTCs are completely equivalent to the path-integral formulation of CTCs developed by Politzer \cite{politzer_path_1994}, as shown in \cite{lloyd_quantum_2011}.

Importantly, while these two prescriptions are distinct, they each resolve time-travel paradoxes in their own way. P-CTCs provide unique resolutions, whereas D-CTCs require an additional condition (see \cite{ralph_information_2010,ralph_reply_2011,pienaar_quantum_2011,ralph_relativistic_2012,allen_treating_2014,dong_ralphs_2017}) to pick out a unique solution. Since both are treatments of interacting quantum systems near CTCs, however, we observe nonunitarity in the evolutions of CR systems in both, which in turn means that the output states are nonlinear functions of the input states. In the case of P-CTCs, the output states remain pure (if initially so), but become unnormalised (thereby requiring renormalisation), and such an effect jeopardises the usual probabilistic interpretation of quantum states. On the other hand, D-CTCs produce entropy through mixing of the input states. Additionally, another important difference between the two models is that for D-CTCs, there is no restriction on the initial data, while with P-CTCs, the renormalisation leads to limits on the initial data as a function of the future.

Nonlinearity is a highly nontrivial issue as its presence fundamentally changes the structure of the theory. This is because the standard proofs of many key theorems in quantum mechanics, which include the no-signalling, no-cloning, and indistinguishability of nonorthogonal states theorems, depend on its linearity. Therefore, given the nonlinearity and nonunitarity of D-CTCs and P-CTCs, it is unsurprising that they have some very interesting applications. Specifically, both D-CTCs \cite{brun_localized_2009} and P-CTCs \cite{brun_perfect_2012} have uses in distinguishing nonorthogonal quantum states, while D-CTCs \cite{ahn_quantum-state_2013,brun_quantum_2013} can clone arbitrary states and P-CTCs can both signal to the past \cite{ralph_relativistic_2012,bub_quantum_2014,ghosh_quantum_2018} and delete arbitrary states. Consequences such as these, in addition to the possibility that either theory accurately describes time travel, undeniably motivates further study into D-CTCs and P-CTCs.

\subsection{Quantum billiard ball paradox}

\begin{figure}[b]
	\vspace{-0.35cm}
	\hspace*{-0.55cm}
	\includegraphics{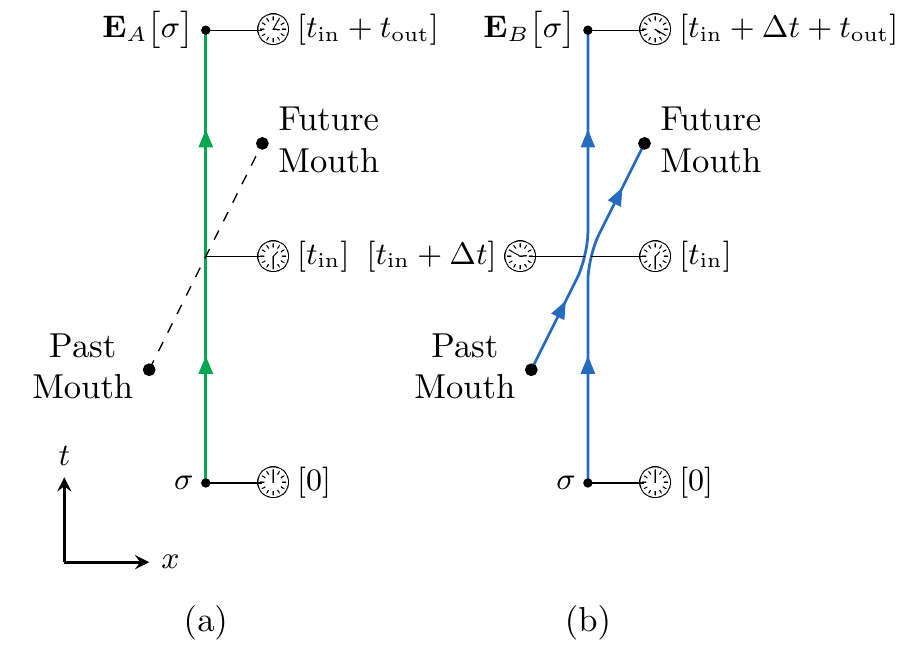}
	\caption[Billiard clocks]{\label{fig:billiard_clocks}Based on the simplified billiard-ball paradox in Fig.~\ref{fig:billiard_balls}, these diagrams in $(1+1)$-dimensional spacetime visualise the basic idea of the model under study in the reference frame of the clock. (a) The history where the clock travels through the CTC-wormhole region with no interaction; (b) the history of a clock which elastically interacts with itself. The times as measured by the clocks at specific points along the histories are given in brackets next to their analogue clock symbols. These clocks begin at time $\TimeGeneral=0$ and, due to their initial velocity, measure the proper times $\TimeGeneral_\EvolutionA=\TimeInbound+\TimeOutbound$ and $\TimeGeneral_\EvolutionB=\TimeInbound+\DeltaTime+\TimeOutbound = \TimeGeneral_\EvolutionA + \DeltaTime$ for evolutions $\EvolutionA$ and $\EvolutionB$, respectively. This means that we denote the duration of the segment on the CTC in evolution $\EvolutionB$ to be $\DeltaTime$, while the inbound and outbound times to and from the wormhole axis (dashed line) are $\TimeInbound$ and $\TimeOutbound$, respectively. It is important to note that the final position bears no information about the path the ball has taken.}
\end{figure}

By its very nature, classical physics simply cannot make sense of solution multiplicity. Despite this, quantum treatments of specific instances of classical multiplicity in time-travel paradoxes have not actually been investigated in the literature. Prompted by this notable absence of such studies, in this paper we present a quantum model of the billiard-ball paradox in which:
\begin{enumerate}[label=(\roman*),itemsep=0.015cm]
	\item We consider a simplified $(1+1)$-dimensional version of the classical problem in which, by virtue of the spacetime dimensionality and geometry of the CTC-wormhole, there are \emph{only} two possible classical paths (see Fig.~\ref{fig:billiard_clocks}).
	\item This problem is mapped to a quantum circuit representation and we use a vacuum state to allow the particles to be present or absent from particular paths.
	\item An internal degree of freedom that operates like a clock is included into the particle's quantum description, meaning that a measurement of the clock's final state can reveal its path's proper time and therefore which of the two classical alternatives was realised.
\end{enumerate}
	
We emphasise the fact that the scenario being effectively $(1+1)$ [point (i)] is not a characteristic specified by Friedman \emph{et al.}, but is part of our modelling of the paradox. In addition, our model is based upon two distinct mechanisms. The first of these [point (ii)] involves the introduction of a ``vacuum'' state. By considering such a state to represent the absence of a billiard ball, we can use it to allow the associated clock to either travel unperturbed (if there is nothing, i.e., a vacuum, in the CTC) or else be scattered into the CTC (if the billiard ball's future self is trapped inside the CTC). The second mechanism [point (iii)] of our model consists of an internal degree of freedom which is incorporated into the billiard ball's quantum description. Using this modification, we can measure the proper time of the model's distinct classical evolutions, which is to say that the billiard ball effectively functions like a clock. This allows us to operationally extract which-way information, i.e., determine which path the billiard ball experienced. This is necessary to give operational meaning to the probabilities associated with the two classical paths, which are indistinguishable by position measurements alone.

It is important to reiterate that the presence of a clock makes the two evolutions of the simple billiard-ball paradox distinguishable. In the Friedman \emph{et al.}~conjecture \cite{friedman_cauchy_1990}, the use of a semiclassical sum-over-histories approach is only meaningful when the evolutions $\EvolutionA$ and $\EvolutionB$ are treated distinctly \emph{a priori}. This enables one to assign evolution probabilities without the need for conventional which-way information. We, however, stress that Friedman \emph{et al.}~do not provide an operational prescription on how to associate paths with probability, which is the reason why we employ a quantum clock.

One of the open questions that we investigate using our model is whether a quantum description does indeed address the issue of evolution indeterminism. In our findings, we observe a parametrisation of the solutions in the D-CTCs description. Although usually interpreted as solution multiplicity, we discuss how this arises naturally as a choice of ``initial state'' in the CTC. Alternatively, in P-CTCs, only one quantum state is offered by the prescription as a solution to the paradox. Rather simply, the state takes the form of a pure superposition of the clock having evolved and not evolved on the CTC during its history through the time machine region (with the possible addition of a vacuum component depending on the specification of the input state). A focus of these P-CTCs findings is how this superposition lends credence to the semiclassical billiard-ball conjecture of Friedman \emph{et al.}~in Ref.~\cite{friedman_cauchy_1990}. We also show how the well-known P-CTC constraints on the initial data (where such restrictions depend on the future of the time-travelling state) limit the actions that one is able to perform in the billiard-ball interaction.

We begin with a description of our quantum billiard-ball paradox model in Sec.~\ref{sec:model} (which includes a specification of our clock state in Sec.~\ref{sec:clocks}). General analytical results of the model are given in Sec.~\ref{sec:results}, while discussions ensue in Sec.~\ref{sec:discussion}. Concluding remarks are lastly made in Sec.~\ref{sec:conclusion}.

\section{\label{sec:model}Model}

Our quantum billiard ball is modelled as a qudit moving in a $(1+1)$-dimensional spacetime in which wormhole mouths appear and disappear in such a way as to allow only two different paths (see Fig.~\ref{fig:billiard_clocks}). We use the qudit degree of freedom to encode a clock on the particle as described in the subsequent section. It is important to note that the particle's position degree of freedom starts in a semiclassical wave packet, with negligible probability that it falls into the wormhole by its own free evolution. This is a significant point, as it allows us to justify both dropping position from the particle's description and the consideration of only the two trajectories $\EvolutionA$ and $\EvolutionB$. It also connects with the WKB approximation mentioned by Friedman \emph{et al.}, who propose to apply the sum-over-histories only to the semiclassical trajectories (thus excluding trajectories that fall into the wormhole without collision).

For simplicity and rigour, we work in $(1+1)$ dimensions, which means the only interaction that can occur between the billiard-ball clock's future and past selves is a complete exchange of momentum. This is a useful characteristic, as a quantum \textgate{SWAP} gate between two qudit channels perfectly replicates this action in the quantum circuit framework. To allow the billiard-ball paradox for our clock to function as intended, we will also require the CTC-wormhole to be dynamical. This means that the mouths of the wormhole appear and disappear at exactly the right points in the spacetime to allow our clock to evolve in two distinct ways within the CTC region. The resulting noncollisional evolution $\EvolutionA$ is trivial; the particle remains stationary as the past wormhole mouth appears and disappears behind it, while the future mouth later appears and disappears in front of the particle. Nothing emerges from or enters the wormhole. On the other hand, in the collisional evolution $\EvolutionB$, a moving particle emerges from the appearing past mouth and strikes the stationary particle, transferring all its momentum to it. The struck particle then travels into the future mouth that appears in front of it, while the other particle becomes stationary and remains so in place of its collision partner.

While these trajectories are classical, the particle can be in superpositions of being absent or present on either path. We stress that the particle's position degree of freedom factors out of the evolution because the two classical trajectories coincide in final position and velocity. Therefore, the only relevant degrees of freedom are the internal states of the clock, plus the presence or absence of the clocks.

It is also important to note that the incoming past and outgoing future momenta of the billiard-ball clock must exactly match. Assuming a perfectly elastic collision, the combination of the dynamical wormhole and the law of conservation of momentum collectively constrains the velocity of the billiard-ball clock on the CTC to a single value (determined by both the geometry of the wormhole and mass of the billiard ball). As a result, there is no need to restrict the paths of the clock (e.g., with waveguides) or its initial velocity because the two distinct, desired histories of our model arise naturally as the \emph{only} two paths through the spacetime. With this, we can represent the two evolutions of our clock in its reference frame through the CTC region as per Fig.~\ref{fig:billiard_clocks}, where we denote the proper travel time of the clock through the wormhole to be exactly $\DeltaTime$.

\subsection{\label{sec:clocks}Quantum clocks}

We introduce an internal qudit degree of freedom to the billiard ball that through its evolution, tells us how much time has passed (in its reference frame). In other words, we endow the billiard ball with a ``clock.'' For our purposes, this quantum modification serves to characterise the difference in proper time between evolutions $\EvolutionA$ and $\EvolutionB$. Specifically, if this internal degree of freedom is observable in some manner (e.g., by looking at the ball's ``wristwatch'' upon its emergence from the CTC region), then, upon measurement, we can deduce which evolution the particle undertook. Such a modelling of an internal clock to track a quantum particle's proper time was introduced in \cite{zych_quantum_2011}.

For our analysis, we employ the $N$-level, pure, equiprobabilistic quantum clock state
\begin{equation}
	\aket{\qclock} = \frac{1}{\sqrt{\NumberLevels}}\sum_{n=1}^{\NumberLevels}\aket{n}, \label{eq:clock_pure}
\end{equation}
where the number states $\{\aket{n}\}_{n=1}^\NumberLevels$, which obey orthonormality via the condition $\abraket{m}{n} = \delta_{nm}$, collectively form a basis for the $N$-dimensional Hilbert space in which the clock resides. We then define the clock Hamiltonian as
\begin{equation}
	\op{H}_\mathrm{c} = \sum_{n=1}^{\NumberLevels}E_n\aket{n}\abra{n} \label{eq:clock_Hamiltonian}
\end{equation}
where $E_n$ is the energy of the $n$th eigenstate $\aket{n}$. Time evolution of a state over the times $\TimeGeneralFirst\rightarrow\TimeGeneralSecond$ is then generated by the time evolution unitary, which in its standard form may be written
\begin{equation}
	\op{\timeunitary}(\TimeGeneralSecond;\TimeGeneralFirst) \equiv \op{\timeunitary}(\TimeGeneralSecond-\TimeGeneralFirst) = \e^{-\eye\op{H}_\mathrm{c}(\TimeGeneralSecond-\TimeGeneralFirst)/\hbar}. \label{eq:time_evolution}
\end{equation}
Equation (\ref{eq:clock_Hamiltonian}) allows us to express an evolved state $\aket{\qclock(\Time)}$ in terms of its constituent basis states $\{\aket{n}\}_{n=1}^\NumberLevels$ as
\begin{equation}
	\aket{\qclock(\Time)} = \frac{1}{\sqrt{\NumberLevels}}\sum_{n=1}^{\NumberLevels}\e^{-\eye E_n \Time/\hbar}\aket{n}. \label{eq:clock_evolved}
\end{equation}
The overlap $\abraket{\qclock(\Time)}{\qclock(\Time+\DeltaTime)}$ between two states at different stages of time evolution then quantifies the indistinguishability (i.e., coherence) between the two clock states. Now, given our freedom of choice in specifying the energies without losing generality or functionality of the clock, we may choose them to be equally spaced such that the $n$th energy $E_n$ can be expressed in terms of the ground state $E_1$ and the spacing $\DeltaEnergy = E_n - E_{n-1}$ as
\begin{equation}
	E_n = E_1 + (n-1)\DeltaEnergy. \label{eq:energies}
\end{equation}
From this, we introduce the orthogonalisation time of our $\NumberLevels$-level clock, defined as
\begin{equation}
	\OrthogonalisationTime = \frac{2\pi\hbar}{\NumberLevels\DeltaEnergy}, \label{eq:orthogonalisation_time}
\end{equation}
which represents the minimal time needed for a clock to evolve [under the now equally spaced Hamiltonian (\ref{eq:clock_Hamiltonian})] into an orthogonal one. A system with finite $\OrthogonalisationTime$ can be thought of as a clock which ``ticks'' at a rate proportional to $\OrthogonalisationTime^{-1}$. Using the energy level relation (\ref{eq:energies}) then allows us to write the clock overlap in terms of the orthogonalisation time (\ref{eq:orthogonalisation_time}) as
\begin{align}
	\abraket{\qclock(\Time)}{\qclock(\Time+\DeltaTime)} = \frac{\e^{-\eye E_1\DeltaTime/\hbar}}{\NumberLevels} \sum_{n=1}^{\NumberLevels} \exp\left[-2\pi\eye\frac{n-1}{\NumberLevels}\frac{\DeltaTime}{\OrthogonalisationTime}\right]. \label{eq:clock_overlap_orthogonalisation}
\end{align}
In the case that the evolution time difference is equal to the orthogonalisation time, i.e, $\DeltaTime = \OrthogonalisationTime$, then one can show that the clock overlap vanishes, i.e.,
\begin{equation}
	\left.\abraket{\qclock(\Time)}{\qclock(\Time+\DeltaTime)}\right|_{\DeltaTime = \OrthogonalisationTime} = 0.
\end{equation}
The case of $\DeltaTime = \OrthogonalisationTime$ can thus be interpreted as when the clock's ``resolution'' exactly matches the time difference between the relevant states. This is the key mechanism with which one can investigate temporal differences between the multiple trajectories (such as the two paths in the simple billiard-ball paradox) that are generated via the indeterminism present in spacetimes containing CTCs.

As a final point, note that the simple forms of the clock state and its associated Hamiltonian are only convenient choices\textemdash the results we obtain in this paper are completely independent of them (provided the initial clock is not an energy eigenstate, but is a superposition of different energies). In particular, given that any quantum system can be decomposed into an orthogonal energy basis, our use of such a basis poses no restrictions on our findings. The use of uniform $1/\sqrt{\NumberLevels}$ weights in the initial state also merely provides simplicity compared to that of general amplitudes, $\{c_n\}_{n=1}^{\NumberLevels}$. In the same vein, the fact that the energy levels are chosen to be equally spaced means that the orthogonalisation time is the same between orthogonal states.

\subsection{Quantum circuit model of the billiard ball paradox}

\begin{figure}
	\includegraphics{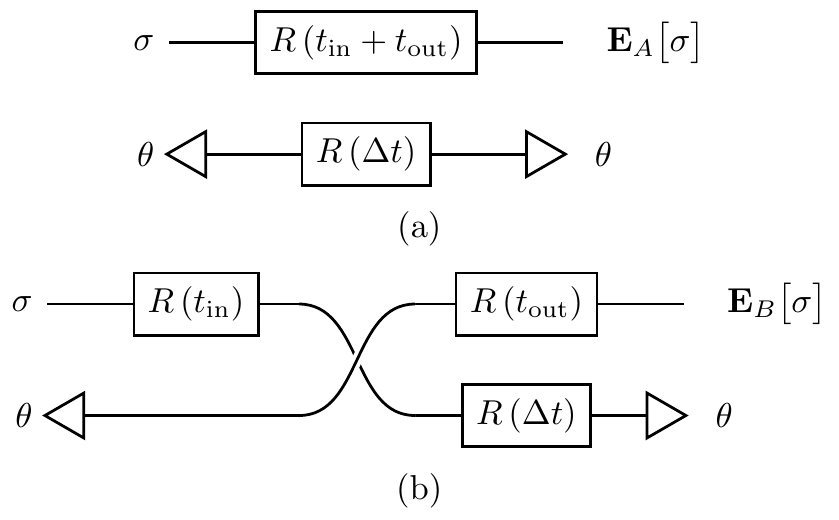}
	\caption[Circuit model]{\label{fig:circuit_model}Quantum circuit models of the two trajectories under study. (a) The evolution $\EvolutionA$ in Fig.~\ref{fig:billiard_clocks}(a); (b) the evolution $\EvolutionB$ in Fig.~\ref{fig:billiard_clocks}(b).}
\end{figure}

The model which we study in this paper involves the evolution of the quantum clock state,
\begin{equation}
	\op{\CRState} = \aket{\qclock}\abra{\qclock}, \label{eq:clock_density}
\end{equation}
along the trajectories illustrated in Fig.~\ref{fig:billiard_clocks}. Given these trajectories, it is easy to construct quantum circuit formulations for the clock states. Such circuits appear in Fig.~\ref{fig:circuit_model}. Here, the elastic self-collision of the clock between its past and future selves may be represented by the \textgate{SWAP} gate,
\begin{equation}
	\op{\Swap} = \sum_{i,j=1}^{\NumberLevels}{\aket{i}\abra{j}}_\CR\otimes{\aket{j}\abra{i}}_\CV. \label{eq:SWAP}
\end{equation}
In effect, this exchange of the CV (trapped CTC) and CR (incoming system) quantum states mimics the momentum-exchange collision interaction between the billiard ball's past and future selves [which necessarily occurs in $(1+1)$ dimensions]. Next, in order to combine the two separate subcircuits of Fig.~\ref{fig:circuit_model} into a single model, we introduce a vacuum state $\aket{0}$ (orthonormal to the existing number states) into both the CR and CV Hilbert spaces. We mandate that the vacuum and clock cannot ``collide'' with each other, and this effect will be introduced by modifying the \textgate{SWAP} gate to the form
\begin{align}
	\op{\vac{\Swap}} &= \op{\vac{\Identity}}_\CR\otimes\aket{0}\abra{0}_\CV + \aket{0}\abra{0}_\CR\otimes\op{\vac{\Identity}}_\CV \nonumber\\
	&\quad - \aket{0}\abra{0}_\CR\otimes\aket{0}\abra{0}_\CV + \op{\Swap} \label{eq:SWAP_vacuum}
\end{align}
where $\op{\vac{\Identity}} = \op{\Identity} + \aket{0}\abra{0}$ is the identity matrix in a vacuum-inclusive Hilbert space. This altered \textgate{SWAP} simply excludes the vacuum from swapping with the nonvacuous states. In this interpretation, the state $\aket{0}$ represents the physical absence of a clock $\aket{\qclock}$, and hence the ``vacuum'' nomenclature. In the associated extended Hilbert space, we write the vacuum-inclusive time evolution unitary as
\begin{equation}
	\op{\vac{\Rotation}}(\TimeGeneralSecond - \TimeGeneralFirst) = \aket{0}\abra{0} + \op{\Rotation}(\TimeGeneralSecond - \TimeGeneralFirst),
\end{equation}
where, for simplicity, we assigned the vacuum with a vanishing energy such that its time evolution phase coefficient is unity. Next, the pure input clock state (\ref{eq:clock_pure}) is modified to include the vacuum,
\begin{equation}
	\aket{\vac{\qclock}} = \sqrt{\Vacuum}\aket{0} + \sqrt{1-\Vacuum}\aket{\qclock}, \qquad 0\leq\Vacuum\leq 1, \label{eq:clock_vacuum_ket}
\end{equation}
which obeys the normalisation condition $\abraket{\vac{\qclock}}{\vac{\qclock}} = 1$ and can be expressed in a density form as
\begin{equation}
	\op{\vac{\CRState}} = \aket{\vac{\qclock}}\abra{\vac{\qclock}}. \label{eq:vacuum_clock_density}
\end{equation}
This is simply the vacuum-inclusive version of the clock density (\ref{eq:clock_density}) and is used as input to our circuit.

\begin{figure}
	\includegraphics{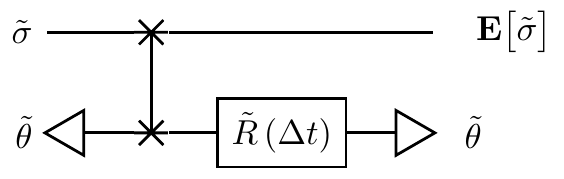}
	\caption[Vacuum model]{\label{fig:circuit_vacuum}The vacuum-based quantum circuit formulation of the billiard-ball paradox (see Fig.~\ref{fig:billiard_balls}) under study. The connected crosses between the two channels represent our prescribed vacuum-excluding $\textgate{SWAP}$ gate.}
\end{figure}

Given the results from the preliminary analyses of the circuits in the previous section, we can, without loss of generality, set for simplicity the inbound and outbound times both equal to zero, i.e., $\TimeInbound = \TimeOutbound = 0$, for all subsequent analysis. This means that a clock which does not interact with the CTC will not time evolve, while one which does interact will time evolve (by the CTC ``duration'' $\DeltaTime$). With this in mind, we may depict our billiard-ball paradox circuit model as per Fig.~\ref{fig:circuit_vacuum}, of which the corresponding vacuum-modified unitary is
\begin{equation}
	\op{\vac{\UnitaryVacuum}} = \left[\op{\vac{\Identity}}_\CR\otimes\op{\vac{\Rotation}}_\CV(\DeltaTime)\right]\op{\vac{\Swap}}. \label{eq:unitary_vacuum}
\end{equation}
Under this, the input state may either self-consistently interact with the CTC or it may pass straight through the circuit, thus allowing it to evolve through the circuit in the two paths of Fig.~\ref{fig:billiard_clocks}.

\section{\label{sec:results}Results}

Here we present the general analyses of the model using the $\NumberLevels$-level clocks introduced in Sec.~\ref{sec:clocks}. For supplementary comprehension, visualisations using qubit clocks of the various relevant quantities appear in the Appendix.

\subsection{D-CTCs results}

With the unitary (\ref{eq:unitary_vacuum}) and CR input (\ref{eq:vacuum_clock_density}), CV fixed points present themselves as solutions to the equation
\begin{widetext}
\vspace*{-1.15cm}
\begin{align}
	\op{\vac{\CVState}} = \Vacuum\op{\vac{\Rotation}}\op{\vac{\CVState}}\op{\vac{\Rotation}}^\dagger + \left(1-\Vacuum\right)\left\{\abra{0}\op{\vac{\CVState}}\aket{0} \aket{0}\abra{0} + \left(1 - \abra{0}\op{\vac{\CVState}}\aket{0}\right)\op{\Rotation}\aket{\qclock(0)}\abra{\qclock(0)}\op{\Rotation}^\dagger + \frac{1}{\NumberLevels}\sum_{i,j=1}^{\NumberLevels}\op{\vac{\Rotation}}\Bigl[\abra{i}\op{\vac{\CVState}}\aket{0}\aket{j}\abra{0} + \abra{0}\op{\vac{\CVState}}\aket{j}\aket{0}\abra{i}\Bigr]\op{\vac{\Rotation}}^\dagger\right\}. \label{eq:vacuum_D-CTCs_fixed}
\end{align}
When $\Vacuum = 1$, the input state is purely vacuous, and so any cross-channel interactions are prevented by the nature of the \textgate{SWAP} gate (\ref{eq:SWAP_vacuum}). Therefore, the trapped state can be any time-independent density matrix, namely, any mixture of energy eigenstates (i.e., diagonal in the energy basis). On the other hand, for $\Vacuum < 1$, the CR and CV channels interact nontrivially, resulting in a stronger constraint. First introducing the evolved clock notation
\begin{align}
	\op{\Rotation}^k(\TimeGeneralSecond - \TimeGeneralFirst)\aket{\qclock} \equiv \aket{\qclock^{(k)}(\TimeGeneralSecond - \TimeGeneralFirst)},
\end{align}
one can verify that the general solution takes the form
\begin{align}
	\map{\DCTCsCVMap}_{\vac{\UnitaryVacuum}}[\op{\vac{\CRState}},\op{\vac{\CVState}}](\DeltaTime,\Vacuum,\Free) =
	\begin{dcases}
		\Free\aket{0}\abra{0} + \left(1 - \Free\right)\op{\varrho}, \quad &\Vacuum=1,\\
		\Free\aket{0}\abra{0} + (1-\Free)\frac{(1-\Vacuum)}{\Vacuum}\sum_{k=1}^{\infty}\Vacuum^k \aket{\qclock^{(k)}(\DeltaTime)} \abra{\qclock^{(k)}(\DeltaTime)}, \quad &\Vacuum<1;
	\end{dcases} \label{eq:vacuum_D-CTCs_CV}
\end{align}
\vspace{-0.2cm}
\end{widetext}
where $0 \leq \Free \leq 1$ is a free parameter and $\op{\varrho}$ is a arbitrary classical mixture in the clock subspace, i.e.,
\begin{equation}
	\op{\varrho} = \sum_{n=1}^{\NumberLevels} c_{n} \aket{n}\abra{n}.
\end{equation}
The parameter $\Free$ can be interpreted as the probability that the CTC contains no clock, and its presence is usually taken to indicate a multiplicity in the set of solutions. An alternative interpretation is that $\Free$ parametrizes some particular properties of the wormhole itself. Note that when $\Vacuum = 1$, the vacuous input state does not interact with the classical CV mixture. Physically, we can interpret the clock subspace portion $\op{\varrho}$ as a preexisting state of the CTC's time evolution, which manifested before we began our theoretical experiment. This is to say that $\op{\varrho}$ is unmeasurable and merely represents some classical configuration of energy that was perhaps captured by the wormhole during its construction. In any case, as the trapped state is not interacted with when $\Vacuum = 1$, we hereafter restrict our attention to the more physically interesting regime of $\Vacuum < 1$ wherein we will retain the freedom of $\Free$ and track its effect through the calculations.

For the measurable part of the solution, observe how it takes the form of a simple mixture of the vacuum and an infinite series of clock states at different evolutionary times. This spectrum of clock CTC ``windings'' is weighted by the input vacuum coefficient $\Vacuum$ such that in the case of $\Vacuum<1$, higher-order windings have lesser probability to be trapped within the CTC.

The D-CTCs output state corresponding to the fixed point (\ref{eq:vacuum_D-CTCs_CV}) is found to be
\begin{widetext}
\vspace{-0.4cm}
\begin{align}
	\map{\DCTCsCRMap}_{\vac{\UnitaryVacuum}}[\op{\vac{\CRState}},\op{\vac{\CVState}}](\DeltaTime,\Vacuum,\Free) &= \Free\aket{\vac{\qclock}(0)}\abra{\vac{\qclock}(0)} + (1-\Free)\frac{(1-\Vacuum)}{\Vacuum}\sum_{k=1}^{\infty}\Vacuum^k\Biggl[\Vacuum\aket{0}\abra{0} + (1-\Vacuum)\aket{\qclock^{(k)}(\DeltaTime)}\abra{\qclock^{(k)}(\DeltaTime)} \nonumber\\
	&\quad + \sqrt{\Vacuum}\sqrt{1-\Vacuum}\frac{\tr[\op{\Rotation}^k(\DeltaTime)]}{\NumberLevels}\aket{0}\abra{\qclock^{(k)}(\DeltaTime)} + \sqrt{\Vacuum}\sqrt{1-\Vacuum}\frac{\tr[\op{\Rotation}^{\dagger k}(\DeltaTime)]}{\NumberLevels}\aket{\qclock^{(k)}(\DeltaTime)}\abra{0}\Biggr]	\label{eq:vacuum_D-CTCs_CR}
\end{align}
\vspace{-0.2cm}
\end{widetext}
where we used
\begin{equation}
	\abraket{\qclock(0)}{\qclock^{(k)}(\DeltaTime)} = \frac{\tr[\op{\Rotation}^k(\DeltaTime)]}{\NumberLevels}.
\end{equation}

Let us now analyse these results. First, it is easy to analytically verify that the state populations are constant with respect to $\DeltaTime$, and in fact take the values
\begin{align}
	\abra{n}\map{\DCTCsCVMap}_{\vac{\UnitaryVacuum}}\aket{n} &=
	\begin{cases}
	\Free, \quad &n=0,\\
	\frac{1}{\NumberLevels}\left(1 - \Free\right), \quad &n>0;
	\end{cases} \label{eq:vacuum_D-CTCs_CV_populations}
\end{align}
\begin{align}
	\abra{n}\map{\DCTCsCRMap}_{\vac{\UnitaryVacuum}}\aket{n} &=
	\begin{cases}
		\Vacuum, \quad &n=0,\\
		\frac{1}{\NumberLevels}\left(1 - \Vacuum\right), \quad &n>0.
	\end{cases} \label{eq:vacuum_D-CTCs_CR_populations}
\end{align}
These results tell us that the parameter $\Free$ directly controls the mixing between vacuous and nonvacuous levels for the CV trapped state, while $\Vacuum$ has exactly the same effect on the CR output state.

Next, we can compute the probabilities of measuring unevolved and orthogonal clocks in both D-CTCs states to be
\begin{widetext}
\begin{subequations}
\vspace*{-1.15cm}
\begin{align}
	\abra{\qclock(0)}\map{\DCTCsCVMap}_{\vac{\UnitaryVacuum}}\aket{\qclock(0)} &= (1-\Free)\frac{(1-\Vacuum)}{\Vacuum}\sum_{k=1}^{\infty}\Vacuum^k\frac{\tr[\op{\Rotation}^k(\DeltaTime)]\tr[\op{\Rotation}^{\dagger k}(\DeltaTime)]}{\NumberLevels^2},\\
	\abra{\qclock(\OrthogonalisationTime)}\map{\DCTCsCVMap}_{\vac{\UnitaryVacuum}}\aket{\qclock(\OrthogonalisationTime)} &= (1-\Free)\frac{(1-\Vacuum)}{\Vacuum}\sum_{k=1}^{\infty}\Vacuum^k\frac{\tr[\op{\Rotation}^\dagger(\OrthogonalisationTime)\op{\Rotation}^k(\DeltaTime)]\tr[\op{\Rotation}^{\dagger k}(\DeltaTime)\op{\Rotation}(\OrthogonalisationTime)]}{\NumberLevels^2};
\end{align}\label{eq:vacuum_D-CTCs_CV_probabilities}
\end{subequations}
\begin{subequations}
\begin{align}
	\abra{\qclock(0)}\map{\DCTCsCRMap}_{\vac{\UnitaryVacuum}}\aket{\qclock(0)} &= \Free +	(1-\Free)\frac{(1-\Vacuum)^2}{\Vacuum}\sum_{k=1}^{\infty}\Vacuum^k\frac{\tr[\op{\Rotation}^k(\DeltaTime)]\tr[\op{\Rotation}^{\dagger k}(\DeltaTime)]}{\NumberLevels^2},\\
	\abra{\qclock(\OrthogonalisationTime)}\map{\DCTCsCRMap}_{\vac{\UnitaryVacuum}}\aket{\qclock(\OrthogonalisationTime)} &= (1-\Free)\frac{(1-\Vacuum)^2}{\Vacuum}\sum_{k=1}^{\infty}\Vacuum^k\frac{\tr[\op{\Rotation}^\dagger(\OrthogonalisationTime)\op{\Rotation}^k(\DeltaTime)]\tr[\op{\Rotation}^{\dagger k}(\DeltaTime)\op{\Rotation}(\OrthogonalisationTime)]}{\NumberLevels^2},
\end{align}\label{eq:vacuum_D-CTCs_CR_probabilities}
\end{subequations}
\vspace{-0.2cm}
\end{widetext}
where we used the fact that
\begin{align}
	0 = \abraket{\qclock(0)}{\qclock(\OrthogonalisationTime)} = \frac{\tr[\op{\Rotation}(\OrthogonalisationTime)]}{\NumberLevels}. \label{eq:orthogonal_trace}
\end{align}
These quantities tell us the relative likelihoods of detecting either type of clock trapped within the CTC or exiting the CTC region. From them, we deduce the relations
\begin{subequations}
\begin{align}
	\abra{\qclock(0)}\map{\DCTCsCRMap}_{\vac{\UnitaryVacuum}}\aket{\qclock(0)} &= \Free + (1-\Vacuum)\abra{\qclock(0)}\map{\DCTCsCVMap}_{\vac{\UnitaryVacuum}}\aket{\qclock(0)},\\
	\abra{\qclock(\OrthogonalisationTime)}\map{\DCTCsCRMap}_{\vac{\UnitaryVacuum}}\aket{\qclock(\OrthogonalisationTime)} &= (1-\Vacuum)\abra{\qclock(\OrthogonalisationTime)}\map{\DCTCsCVMap}_{\vac{\UnitaryVacuum}}\aket{\qclock(\OrthogonalisationTime)},
\end{align}
\end{subequations}
with which we infer that the D-CTCs trapped and output state clock probabilities are related linearly with respect to $\OrthogonalisationTime$.

\subsection{P-CTCs results}

According to the P-CTCs prescription, to find the output state, we first must trace out the CV channel from the unitary (\ref{eq:unitary_vacuum}), which yields
\begin{equation}
	\op{\vac{\UnitaryTraced}} \equiv \tr_\CV[\op{\vac{\UnitaryVacuum}}] = \tr[\op{\vac{\Rotation}}(\DeltaTime)]\aket{0}\abra{0} + \op{\Identity} + \op{\Rotation}(\DeltaTime). \label{eq:vacuum_P-CTCs_operator}
\end{equation}
With this, one can then compute the output in pure state form to be
\begin{widetext}
\vspace{-0.4cm}
\begin{align}
	\map{\PCTCsMap}_{\vac{\UnitaryVacuum}}\MapBracket{\op{\vac{\CRState}}}(\DeltaTime,\Vacuum) = \frac{\op{\vac{\UnitaryTraced}}\op{\vac{\CRState}}\op{\vac{\UnitaryTraced}}^\dagger}{\tr[\op{\vac{\UnitaryTraced}}\op{\vac{\CRState}}\op{\vac{\UnitaryTraced}}^\dagger]} \sim \frac{1}{\sqrt{\abs{\Normalisation}}}\Biggl\{\sqrt{\Vacuum}\,\tr[\op{\vac{\Rotation}}(\DeltaTime)]\aket{0}+ \sqrt{1-\Vacuum}\Bigl[\aket{\qclock(0)}+ \aket{\qclock(\DeltaTime)}\Bigr]\Biggr\} \label{eq:vacuum_P-CTCs_output}
\end{align}
\vspace{-0.2cm}
\end{widetext}
where the normalisation coefficient $\Normalisation$ is given by
\begin{align}
	\Normalisation &= \tr[\op{\vac{\UnitaryTraced}}\op{\vac{\CRState}}\op{\vac{\UnitaryTraced}}^\dagger] \nonumber\\
	&= \Vacuum\,\tr[\op{\vac{\Rotation}}(\DeltaTime)]\tr[\op{\vac{\Rotation}}^\dagger(\DeltaTime)] \nonumber\\
	&\quad + (1-\Vacuum)\Bigl(2 + \frac{\tr[\op{\Rotation}(\DeltaTime)] + \tr[\op{\Rotation}^\dagger(\DeltaTime)]}{\NumberLevels}\Bigr).
\end{align}
The corresponding state populations are
\begin{equation}
	\abra{n}\map{\PCTCsMap}_{\vac{\UnitaryVacuum}}\aket{n} = \frac{1}{\Normalisation}\times
	\begin{cases}
	\Vacuum\,\tr[\op{\vac{\Rotation}}(\DeltaTime)]\tr[\op{\vac{\Rotation}}^\dagger(\DeltaTime)], \quad &n=0,\\
	(1 - \Vacuum)\abs{1 + \Rotation_{nn}(\DeltaTime)}^2, \quad &n>0.
	\end{cases} \label{eq:vacuum_P-CTCs_populations}
\end{equation}
Unlike the D-CTCs populations in Eqs.~(\ref{eq:vacuum_D-CTCs_CV_populations}) and (\ref{eq:vacuum_D-CTCs_CR_populations}), we observe that these expressions are not simply $\DeltaTime$-independent constants in P-CTCs. Lastly, we can compute the clock probabilities to be
\begin{subequations}
\begin{align}
	\abra{\qclock(0)}\map{\PCTCsMap}_{\vac{\UnitaryVacuum}}\aket{\qclock(0)} &= \frac{1}{\Normalisation}(1-\Vacuum)\abs{1 + \frac{\tr[\op{\Rotation}(\DeltaTime)]}{\NumberLevels}}^2,\\
	\abra{\qclock(\OrthogonalisationTime)}\map{\PCTCsMap}_{\vac{\UnitaryVacuum}}\aket{\qclock(\OrthogonalisationTime)} &= \frac{1}{\Normalisation}(1-\Vacuum) \abs{\frac{\tr[\op{\Rotation}^\dagger(\OrthogonalisationTime)\op{\Rotation}(\DeltaTime)]}{\NumberLevels}}^2.
\end{align}\label{eq:vacuum_P-CTCs_probabilities}
\end{subequations}

\section{\label{sec:discussion}Discussion}

Evidently, our quantum circuit model produces nontrivial results in terms of D-CTCs and P-CTCs. Here, we highlight some interesting features of the results, compare the two prescriptions, discuss the effects of the parameters $\Free$, $\Vacuum$, and $\DeltaTime$, and conclude with some general remarks.

The form of the D-CTCs trapped state becomes intuitive in the equivalent circuit picture (ECP) \cite{ralph_information_2010,ralph_reply_2011,pienaar_quantum_2011,ralph_relativistic_2012,dong_ralphs_2017}. In the ECP, one determines the CV state by iterating an initial seed state many times through the circuit until the fixed point is reached. The corresponding CR state can then be found. An obvious choice for the seed state is the noninteracting trapped state when $\Vacuum = 1$ (\ref{eq:vacuum_D-CTCs_CR}). With this, in each iteration through the circuit [i.e., application of the unitary (\ref{eq:unitary_vacuum})], the nonvacuous portion of the CTC state [initially $(1-\Free)\op{\rho}$] effectively ``catches'' an unevolved clock from the input state into the CTC. Any existing nonvacuous states inside the CTC are then multiplied by the probability that the CR state is vacuous, i.e., $\Vacuum$, which corresponds to the chance that the trapped clock(s) will be unable to leave the CTC. Lastly, everything inside the CTC time evolves, i.e., rotates by the time delay $\DeltaTime$. After an infinite number of interactions, we end up with the infinite spectrum of clocks, $(1-\Free)\sum_{k=1}^{\infty}\Vacuum^k\aket{\qclock^{(k)}(\DeltaTime)} \abra{\qclock^{(k)}(\DeltaTime)}$, which after normalisation by $(1-\Vacuum)/\Vacuum$ and the addition of the unchanged vacuum portion $\Free\aket{0}\abra{0}$ leaves us with our measurable CV solution (\ref{eq:vacuum_D-CTCs_CV}).

The D-CTCs CR output state is then simple to interpret. The vacuous CV component allows for input CR clocks to pass through the circuit unaffected, while the spectrum of trapped clocks swaps with the input clock and combines with the input vacuum in the CR output. Note that if the $\tr[\op{\Rotation}^k(\DeltaTime)]/\NumberLevels$ and $\tr[\op{\Rotation}^{\dagger k}(\DeltaTime)]/\NumberLevels$ factors in the output (\ref{eq:vacuum_D-CTCs_CR}) were unity, then the infinite series term would reduce to the vacuum-clock spectrum, $\sum_{k=1}^{\infty}\Vacuum^k\aket{\vac{\qclock}^{(k)}(\OrthogonalisationTime)} \abra{\vac{\qclock}^{(k)}(\OrthogonalisationTime)}$. This, of course, only occurs when $\DeltaTime=p\NumberLevels\OrthogonalisationTime$ ($p\in\mathds{Z}_{>0}$) however, which is the case where the spectrum disappears due to complete revolution of the clock.

Conversely, the P-CTCs output (\ref{eq:vacuum_P-CTCs_output}) is, excluding the single pure vacuum state, a simple superposition of the unevolved and singly evolved clock states. This is in stark contrast to the spectrum of higher-order windings in D-CTCs associated with $0<\Vacuum<1$. It is only in the case that $\Vacuum=0$ where the D-CTCs output reduces to a classical mixture of the unevolved and singly evolved clock superposition, i.e., $\sqrt{\Free}\aket{\qclock(0)} \pm \sqrt{1-\Free}\aket{\qclock(\DeltaTime)}$.

\subsection{Parameters of the circuit}

\subsubsection{Free parameter \texorpdfstring{$\Free$}{𝑔}}

Normally the free parameter $\Free$ in the D-CTC model is seen as an incompleteness of the model, requiring additional assumptions to remove it. An alternative interpretation emerges from our model. In the ECP, it is easy to analytically verify that the seed state
\begin{equation}
	\op{\vac{\CVState}}_{(0)}(\Free) = \Free\aket{0}\abra{0} + (1-\Free)\op{\rho}, \label{eq:vacuum_D-CTCs_seed}
\end{equation}
(where $\op{\rho}$ is an arbitrary normalised density existing in the clock subspace), evolves to the general trapped state (\ref{eq:vacuum_D-CTCs_CV}). In other words, an infinite number of applications of the unitary (\ref{eq:unitary_vacuum}) to this CV state with the input CR state (\ref{eq:clock_vacuum_ket}) converges to the solution (\ref{eq:vacuum_D-CTCs_CV}). This is because the separate components in the end state fixed point are linear transformations of the input components of the seed state, which consequently means that the initial mixture of the seed state (codified by $\Free$) directly controls the form of the CV state.

Thus, as an alternative and perhaps less involved view, one could simply argue that $\Free$ is not a parameter to be fixed by additional constraints on the model itself, but merely characterises some intrinsic attribute(s) of the wormhole-CTC-time machine (such as geometry, dynamics and/or energy [temperature]).

\subsubsection{Vacuum parameter \texorpdfstring{$\Vacuum$}{Ω}}

Unlike the multiplicity parameter $\Free$, the vacuum input amplitude $\sqrt{\Vacuum}$ influences the results of both D-CTCs and P-CTCs. In the former, its presence causes any nonvacuous components of the CV state to loop around the CTC upon interaction with the CR state. However, in P-CTCs, its effect is trivial in that due to the necessary renormalisation, it merely controls the relative amplitude between the vacuum and clock levels in the output state.

Perhaps intuitively, for the extreme case $\Vacuum = 1$, the trivial vacuous input corresponds to likewise vacuous outputs for both D-CTCs and P-CTCs. Less intuitive is the opposite case where $\Vacuum = 0$, in which the input state consists of only the clock (and no vacuum). Here, of importance is the fact that the D-CTC states no longer contain clocks of rotation order higher than $1$. This is due to the fact that by setting $\Vacuum = 0$, we remove the sole mechanism with which clock states \emph{cannot} swap out of the CTC, thereby eliminating any eternally trapped clocks. As a consequence, the D-CTCs CV state becomes simply a classical mixture of the vacuum and the singly evolved clock, which has a direct interpretation as describing a CTC either containing nothing or a clock, respectively. These two classical alternatives then directly specify what occurs within the CTC region. In the former, where there is probability $\Free$ that there is no clock in the CTC, there is an equal probability of the external clock not interacting with the CTC. Hence, the output clock $\aket{\qclock(0)}$ must have followed the noninteracting evolution $\EvolutionA$. Alternatively, in the latter case, a clock was in the CTC with a complement probability $(1-\Free)$. This can only happen if the initial clock was scattered into it, which means that the evolved clock (which was kicked out by its younger self) takes on the form $\aket{\qclock(\DeltaTime)}$ in accordance with the interacting trajectory $\EvolutionB$.

\subsubsection{CTC time delay \texorpdfstring{$\DeltaTime$}{Δt}}

The phase rotation of the CTC, denoted as $\DeltaTime$ and interpreted as the CTC's time delay, yields which-way information and thus distinguishes the two classical histories of the billiard-ball paradox. To further examine the effect which this has on the solutions, we will look at a few important cases.

First, when $\DeltaTime = 0$, one can interpret the CTC as having a vanishingly short length. As a result, the two histories become, at no fault of our clock, indistinguishable in the sense that the clock does not time evolve during its history through the circuit. Conversely, in the case of $\DeltaTime = \OrthogonalisationTime$, we can use (\ref{eq:orthogonal_trace}) to write the solutions as
\begin{widetext}
\vspace{-0.4cm}
\begin{subequations}
	\begin{align}
	\left.\map{\DCTCsCVMap}_{\vac{\Unitary}}\MapBrackets{\op{\vac{\CRState}}}{\op{\vac{\CVState}}}\right|_{\DeltaTime=\OrthogonalisationTime} &= \Free\aket{0}\abra{0} + (1-\Free)\frac{1-\Vacuum}{1-\Vacuum^{\NumberLevels}}\frac{1}{\Vacuum}\sum_{k=1}^{\NumberLevels}\Vacuum^k\aket{\qclock^{(k)}(\OrthogonalisationTime)} \abra{\qclock^{(k)}(\OrthogonalisationTime)}, \\
	\left.\map{\DCTCsCRMap}_{\vac{\Unitary}}\MapBrackets{\op{\vac{\CRState}}}{\op{\vac{\CVState}}}\right|_{\DeltaTime=\OrthogonalisationTime} &= \Free\aket{\vac{\qclock}(0)} \abra{\vac{\qclock}(0)} + (1-\Free)(1-\Vacuum)\left[\frac{\Vacuum}{1-\Vacuum}\aket{0}\abra{0} + \frac{1-\Vacuum}{1-\Vacuum^{\NumberLevels}}\frac{1}{\Vacuum}\sum_{k=1}^{\NumberLevels}\Vacuum^k\aket{\qclock^{(k)}(\OrthogonalisationTime)} \abra{\qclock^{(k)}(\OrthogonalisationTime)}\right];
	\end{align}
\end{subequations}
\begin{equation}
	\left.\map{\PCTCsMap}_{\vac{\Unitary}}\MapBracket{\op{\vac{\CRState}}}\right|_{\DeltaTime=\OrthogonalisationTime} \sim \frac{\sqrt{\Vacuum}\,\aket{0} + \sqrt{1-\Vacuum}\left[\aket{\qclock(0)} + \aket{\qclock(\OrthogonalisationTime)}\right]}{\sqrt{2-\Vacuum}}.
\end{equation}
\end{widetext}
Here, the D-CTCs clock spectrum of the general $\DeltaTime$ case (\ref{eq:vacuum_D-CTCs_CV}), which consists of (possibly) an infinite number of distinctly rotated clocks $\left\{\op{\Rotation}^{k}(\DeltaTime)\aket{\qclock}\right\}_{k=1}^{\infty}$, reduces to the spectrum of $\NumberLevels$ orthogonal clocks $\left\{\op{\Rotation}^{k}(\OrthogonalisationTime)\aket{\qclock}\right\}_{k=1}^{\NumberLevels}$. In regards to the original billiard-ball paradox, we associate the zeroth orthogonal state $\aket{\qclock(0)}=\op{\Rotation}^{0}(\OrthogonalisationTime)\aket{\qclock}$ with evolution $\EvolutionA$, and the first orthogonal clock $\aket{\qclock(\OrthogonalisationTime)}=\op{\Rotation}^{1}(\OrthogonalisationTime)\aket{\qclock}$ with evolution $\EvolutionB$. The existence of higher-order rotations (i.e., clock windings of the CTC) in D-CTCs merely indicates the entrapment of these states inside the CTC by the external (CR) vacuum. These vanish when the input vacuum amplitude likewise vanishes, i.e., $\Vacuum=0$.

The importance of the time delay $\DeltaTime$ and its relationship with the orthogonalisation time $\OrthogonalisationTime$ become apparent in a physical interpretation of the model. By construction, our clocks have an intrinsic time between ``ticks'' which we denote $\OrthogonalisationTime$. Our $N$-level clocks have $N$ such ticks, or times that we are able to perfectly distinguish between via measurement. This is due to the fact that the only clock states which have vanishing mutual overlaps are those which form the mutually orthogonal clock set. The probabilistic interpretation of our model then leads us to conclude that any clock not in this set will exist in a superposition of orthogonal clocks (i.e., those which are in the set). For example, a clock which has evolved in a nonorthogonal manner possesses a time between two adjacent ticks and, consequently, would have a chance to be in either of the two tick states upon measurement.

Cunning construction of the input clock, such that its orthogonalisation time exactly matches the CTC time delay, yields solutions which consist of clock states that are perfectly distinguishable. As such states correspond to the respective number of times the clock passed through the CTC, we are able, due to the clock's configuration, to extract precise which-way information from the output states.

\subsubsection{\label{sec:special}Special case}

Here, we consider perhaps the most enlightening special case of our model. By mandating that we always send in a clock into the circuit ($\Vacuum = 0$) while ensuring that the CTC's time delay exactly matches the clock's orthogonalisation time ($\DeltaTime = \OrthogonalisationTime$), the output states become
\begin{subequations}
	\begin{align}
	\left.\map{\DCTCsCVMap}_{\vac{\Unitary}}\MapBrackets{\op{\vac{\CRState}}}{\op{\vac{\CVState}}}\right|_{\substack{\Vacuum=0\\\DeltaTime=\OrthogonalisationTime}} &= \Free\aket{0}\abra{0} + (1-\Free)\aket{\qclock(\OrthogonalisationTime)}\abra{\qclock(\OrthogonalisationTime)}, \\
	\left.\map{\DCTCsCRMap}_{\vac{\Unitary}}\MapBrackets{\op{\vac{\CRState}}}{\op{\vac{\CVState}}}\right|_{\substack{\Vacuum=0\\\DeltaTime=\OrthogonalisationTime}} &= \Free\aket{\qclock(0)}\abra{\qclock(0)} + (1-\Free)\aket{\qclock(\OrthogonalisationTime)}\abra{\qclock(\OrthogonalisationTime)};
	\end{align}
\end{subequations}
\vspace{-0.5cm}
\begin{equation}
	\left.\map{\PCTCsMap}_{\vac{\Unitary}}\MapBracket{\op{\vac{\CRState}}}\right|_{\substack{\Vacuum=0\\\DeltaTime=\OrthogonalisationTime}} \sim \frac{1}{\sqrt{2}}\Bigl[\aket{\qclock(0)} + \aket{\qclock(\OrthogonalisationTime)}\Bigr].
\end{equation}
Note that the D-CTCs CR output is a classical mixture of the evolved and unevolved clocks, while the P-CTCs output is an equiprobabilistic superposition of the very same clocks. In fact, given that the postselected teleportation model is equivalent to a path-integral formulation, our circuit appears to lend credence to the original conjecture of Friedman \emph{et al.}~(concerning the WKB approximation of the billiard-ball paradox) in Ref.~\cite{friedman_cauchy_1990}.

\subsection{\label{sec:P-CTCs_constraints}P-CTCs constraints on initial data}

One interesting general characteristic of P-CTCs is that it can pose constraints on initial conditions. Due to the renormalisation, such constraints depend on the future of the time-travelling state. Here, we explicitly demonstrate the restrictions associated with the P-CTCs description of our circuit.

Given our $N$-level clock (\ref{eq:clock_pure}), there exist exactly $N$ mutually orthogonal, equally spaced states $\{\aket{\qclock^{(k)}(\OrthogonalisationTime)}\}_{k=0}^{\NumberLevels-1}$ which it can assume. As a consequence, one may express the clock gate (\ref{eq:time_evolution}) in terms of the ground-state energy $E_1$ defined via (\ref{eq:energies}) as
\begin{align}
	\op{\Rotation}(\TimeGeneralSecond - \TimeGeneralFirst) &= \e^{-\eye E_1 (\TimeGeneralSecond - \TimeGeneralFirst)/\hbar}\nonumber\\
	&\quad\times\sum_{n=1}^{\NumberLevels}\exp\left[-2\pi\eye\frac{n-1}{\NumberLevels}\frac{\TimeGeneralSecond - \TimeGeneralFirst}{\OrthogonalisationTime}\right]\aket{n}\abra{n}. \label{eq:rotation_global}
\end{align}
From this, it is easy to conclude that
\begin{equation}
	\op{\Rotation}^\NumberLevels(\OrthogonalisationTime) = \e^{-\eye \NumberLevels E_1 \OrthogonalisationTime/\hbar}\op{\Identity} \label{eq:rotation_orthogonalisation},
\end{equation}
which simply indicates that $N$ orthogonal rotations of the clock bring it back to its initial state (up to the global phase $\e^{-\eye \NumberLevels E_1 \OrthogonalisationTime/\hbar}$). This, therefore, means that a clock $\aket{\qclock}$ which orthogonalises $p\in\mathds{Z}_{>0}$ times accumulates $p$ such phases. With this, the judicious choice of our clock's orthogonalisation time $\OrthogonalisationTime$, such that it is related to the CTC time delay $\DeltaTime$ by
\begin{equation}
	\DeltaTime = p \NumberLevels \OrthogonalisationTime, \label{eq:orthogonalisation_time_multiple}
\end{equation}
means that a clock which completes one journey on the CTC would orthogonally evolve $p$ times. In such a case, the P-CTCs reduced operator (\ref{eq:vacuum_P-CTCs_operator}) becomes
\begin{align}
	\left.\op{\vac{\UnitaryTraced}}\right|_{\DeltaTime = p \NumberLevels \OrthogonalisationTime} &= \left(1 + \NumberLevels \e^{-\eye p \NumberLevels E_1 \OrthogonalisationTime/\hbar}\right)\aket{0}\abra{0} \nonumber\\
	&\quad + \left(1 + \e^{-\eye p \NumberLevels E_1 \OrthogonalisationTime/\hbar}\right)\op{\Identity}. \label{eq:vacuum_P-CTCs_operator_orthogonalisation}
\end{align}
Additionally, if we construct our clock so that its ground-state energy $E_1$ satisfies
\begin{equation}
	E_1 = \frac{\pi\hbar}{p \NumberLevels \OrthogonalisationTime}(1 + 2q),\qquad q\in\mathds{Z}_{\geq 0}, \label{eq:vacuum_P-CTCs_energy_vanish}
\end{equation}
then the operator (\ref{eq:vacuum_P-CTCs_operator_orthogonalisation}) further reduces to
\begin{align}
	\op{\vac{\UnitaryTraced}}' = \left.\op{\vac{\UnitaryTraced}}\right|_{\substack{\DeltaTime = p \NumberLevels \OrthogonalisationTime\\ E_1 = \frac{\pi\hbar}{p \NumberLevels \OrthogonalisationTime}(1 + 2q)}} = \left(1 - \NumberLevels\right)\aket{0}\abra{0}. \label{eq:vacuum_P-CTCs_operator_constrained}
\end{align}
After renormalisation, the corresponding P-CTCs output is then, of course, simply the vacuum density, i.e., $\aket{0}\abra{0}$. But where did our clock go?

To answer this question, we can use the entangled state
\begin{equation}
	\aket{\psi}_{\text{rec},\CR} = \frac{1}{\sqrt{2}}\Bigl[\aket{0}_\text{rec}\otimes\aket{0}_\CR + \aket{\qclock(0)}_\text{rec}\otimes\aket{\qclock(0)}_\CR\Bigr] \label{eq:vacuum_P-CTCs_entangled}
\end{equation}
as input to our circuit. Here, the state existing in the first Hilbert space may be interpreted as a record of what we send into the second, CR channel. With this entanglement, we find the P-CTCs unnormalised pure state output, corresponding to our circuit's ordinary operator $\op{\vac{\Identity}}_\text{rec}\otimes\op{\vac{\UnitaryTraced}}_\CR$ from (\ref{eq:vacuum_P-CTCs_operator}), to be
\begin{widetext}
\vspace{-0.3cm}
\begin{align}
	\left(\op{\vac{\Identity}}_\text{rec}\otimes\op{\vac{\UnitaryTraced}}_\CR\right)\aket{\psi}_{\text{rec},\CR} \propto \sqrt{\Vacuum}\,\tr[\op{\vac{\Rotation}}(\DeltaTime)]\,\aket{0}_\text{rec}\otimes\aket{0}_\CR + \sqrt{1-\Vacuum}\,\aket{\qclock(0)}_\text{rec}\otimes\Bigl[\aket{\qclock(0)}_\CR + \aket{\qclock(\DeltaTime)}_\CR\Bigr].
\end{align}
\vspace{-0.2cm}
\end{widetext}
The persistence of the entanglement means that when the record indicates that we did not send in a clock, we will not observe a clock in the CR output. Conversely, when we definitely did send in a clock, we must observe a clock exiting the CTC region, which is exactly what one would expect intuitively. However, under the conditions (\ref{eq:orthogonalisation_time_multiple}) and (\ref{eq:vacuum_P-CTCs_energy_vanish}) which yield the operator (\ref{eq:vacuum_P-CTCs_operator_constrained}), the entanglement of our input state (\ref{eq:vacuum_P-CTCs_entangled}) breaks,
\begin{align}
	\left(\op{\vac{\Identity}}_\text{rec}\otimes\op{\vac{\UnitaryTraced}}'_\CR\right)\aket{\psi}_{\text{rec},\CR} &\propto \aket{0}_\text{rec}\otimes\aket{0}_\CR.
\end{align}
The mere presence of the interaction with the CTC in the future implies that the record channel will show that no clock was ever prepared, even though the initial state (\ref{eq:vacuum_P-CTCs_entangled}) included the possibility that a clock was there (which would show up in the record if the future interaction with the CTC was avoided).

Given the path-integral correspondence of P-CTCs, the conditions (\ref{eq:orthogonalisation_time_multiple}) and (\ref{eq:vacuum_P-CTCs_energy_vanish}) collectively form a case in which no clocks can evolve through the circuit due to destructive interference in the path integral. This is to say that P-CTCs suppress all evolutions of nonvacuous states as a result of the particular future evolution of clock states in our model, thereby posing constraints on the initial state. This issue highlights the well-known problem of antichronological and superluminal influence with P-CTCs \cite{ralph_relativistic_2012,bub_quantum_2014,ghosh_quantum_2018}. In contrast, the initial data for D-CTCs are never affected by the presence or absence of the CTC in the future.

\section{\label{sec:conclusion}Conclusion}

In this paper, we presented a quantum circuit formulation of a $(1+1)$-dimensional version of the billiard-ball paradox. Our model is based on two important mechanisms. The first, which involves the incorporation of an internal degree of freedom into the billiard ball, allows it to effectively function like a clock. In practice, this meant that we were able to measure the proper time of the distinct classical evolutions $\EvolutionA$ and $\EvolutionB$, from which we could extract which-way information (distinguishability) regarding these trajectories. The second mechanism is that of the vacuum state, which allowed the clock to either travel unperturbed (if there is nothing, i.e., a vacuum, in the CTC) or be scattered into the CTC (if there is the external clock's trapped future self already inside).

We found that the parametrisation solution arises in the D-CTC prescription, and it was discussed how this resembles the classical solution multiplicity. We also argue how this parametrisation can be naturally interpreted in terms of a choice of ``initial state'' (parametrised by $\Free$) in the D-CTC. On the other hand, the P-CTC model presents only one quantum state as a solution, which takes the form of a pure superposition of the clock having evolved and not evolved through the CTC (with the possible addition of a vacuum component if such a state was initially sent in with the clock). This therefore reproduces the intuition expressed by Friedman \emph{et al.}~in their seminal paper \cite{friedman_cauchy_1990} (discussed in the latter part of Sec.~\ref{sec:introduction} and in Sec.~\ref{sec:special}).

Our results demonstrated that when one prescribes the existence of a vacuum, the D-CTCs and P-CTCs output states become the aforementioned mixture and superposition, respectively, of the histories which do and do not interact with the CTC. These outcomes thus support the notion that if the world operates in accordance with D-CTCs, then the classical picture one gets works like multiple universes; otherwise, if it operates as per P-CTCs, then there are constraints on the actions which one is able to perform.

By basing our quantum circuit model on the classical problem, we assumed that any external degree(s) of freedom relevant to the propagation of the billiard particle do not play a role and can therefore be neglected. Further work could include determining if this is an entirely well-founded postulate, particularly in the context of continuous degrees of freedom. Nevertheless, the framework of incorporating quantum clocks into classical particles that evolve along wholly classical trajectories allowed us to rigorously explore the billiard-ball paradox. Albeit simple, its ability to allow one to distinguish between multiple histories through a chronology-violating region is a powerful function of the methodology, and one could use it to very easily study other time-travel paradoxes.

\begin{acknowledgments}
We wish to thank Magdalena Zych for motivating discussion. This research was supported by the Australian Research Council (ARC) under the Centre of Excellence for Quantum Computation and Communication Technology (Project No.~CE170100012). F.C.~acknowledges support through an Australian Research Council Discovery Early Career Researcher Award (Grant No.~DE170100712). We acknowledge the traditional owners of the land on which the University of Queensland is situated, the Turrbal and Jagera people.
\end{acknowledgments}

\appendix*
\section{\texorpdfstring{$2$}{2}-\uppercase{level qubit clock}}

To supplement the results of the main text, we can employ numerical methods to generate visualisations of the various quantities. For simplicity, these plots can be produced by employing qubit (i.e., $\NumberLevels=2$ dimensions) quantum clock states,
\begin{equation}
	\aket{\qclock(0)} = \frac{1}{\sqrt{2}}\Bigl(\aket{1}+\aket{2}\Bigr), \label{eq:clock_qubit}
\end{equation}
where $\aket{1}$ and $\aket{2}$ are the ground and excited states, respectively. By Eqs.~(\ref{eq:energies}) and (\ref{eq:orthogonalisation_time}), the associated clock gate may be written as
\begin{equation}
	\op{\Rotation}(\DeltaTime) = \e^{-\eye E_1\Delta\TimeGeneral/\hbar}\Bigl(\aket{1}\abra{1} + \e^{-\eye \pi \DeltaTime/\OrthogonalisationTime}\aket{2}\abra{2}\Bigr),
\end{equation}
so that the orthogonalisation time $\DeltaTime = \OrthogonalisationTime$ transforms the initial clock qubit (\ref{eq:clock_qubit}) into the orthogonal state
\begin{equation}
	\aket{\qclock(\OrthogonalisationTime)} = \op{\Rotation}(\OrthogonalisationTime)\aket{\qclock(0)} = \frac{\e^{-\eye E_1\OrthogonalisationTime/\hbar}}{\sqrt{2}}\Bigl(\aket{1}-\aket{2}\Bigr) \label{eq:clock_qubit_orthogonal}.
\end{equation}
Importantly, note that qubit clocks only have two mutually orthogonal states, which means that multiple loops of the CTC [such as those described by the clock spectrums of (\ref{eq:vacuum_D-CTCs_CV}) and (\ref{eq:vacuum_D-CTCs_CR})] cannot be captured by such clocks.

The quantities of interest which we will visualise include the energy level populations (of the $\aket{0},\aket{1}$, and $\aket{2}$ states) and clock probabilities (both the unevolved $\aket{\qclock(0)}$ and orthogonal $\aket{\qclock(\OrthogonalisationTime)}$ clocks). From these diagrams, we can discern the behaviours of the vacuum model in terms of both D-CTCs and P-CTCs. Note that all of our D-CTCs plots will use $\Free=\frac{1}{3}$ in accordance with the rule of maximal entropy.

\subsection{D-CTCs qubit visualisations}

\begin{figure}[b]
	\includegraphics[scale=1.05]{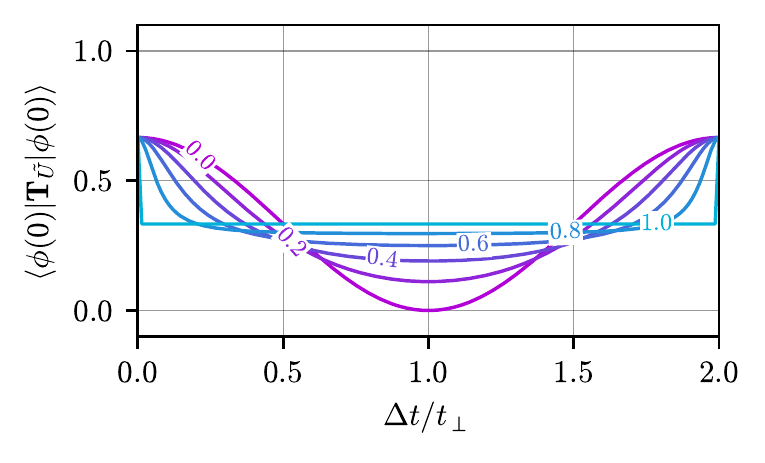}
	\includegraphics[scale=1.05]{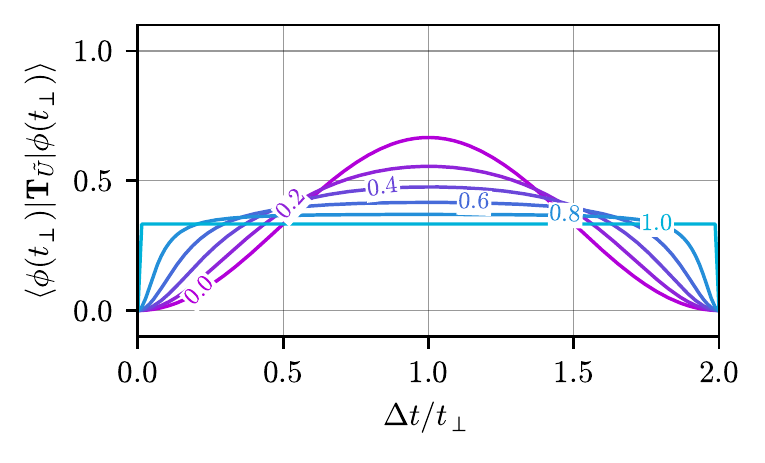}
	\vspace{-0.35cm}
	\caption{\label{fig:vacuum_D-CTCs_CV_probabilities}D-CTCs CV clock probabilities (unitless) for the vacuum model plotted against the dimensionless time ratio. The CTC multiplicity parameter was taken to be $\Free=\frac{1}{3}$, and the lines correspond to varying values of $\sqrt{\Vacuum}$.}
\end{figure}

\begin{figure}
	\includegraphics[scale=1.05]{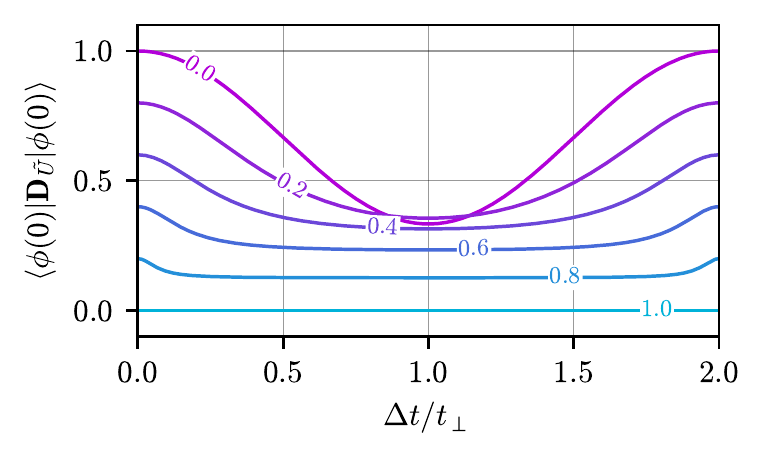}
	\includegraphics[scale=1.05]{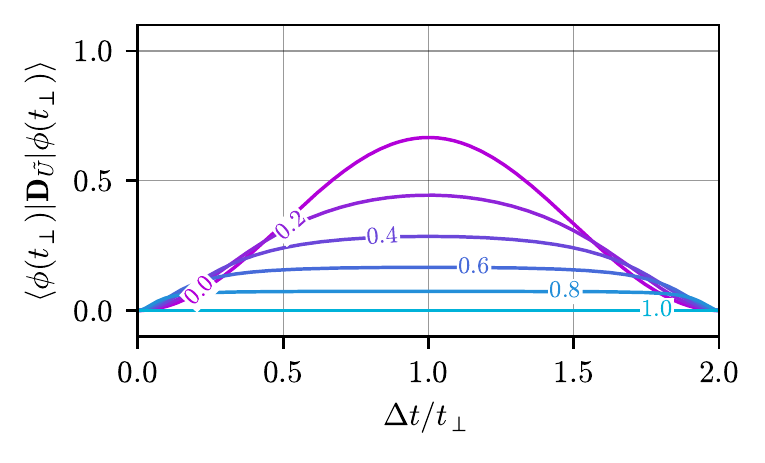}
	\vspace{-0.35cm}
	\caption{\label{fig:vacuum_D-CTCs_CR_probabilities}D-CTCs CR clock probabilities (unitless) for the vacuum model plotted against the dimensionless time ratio. The CTC multiplicity parameter was taken to be $\Free=\frac{1}{3}$, and the lines correspond to varying values of $\sqrt{\Vacuum}$.}
\end{figure}

By employing qubit clocks, it is easy to plot the trapped (\ref{eq:vacuum_D-CTCs_CV_probabilities}) and output (\ref{eq:vacuum_D-CTCs_CR_probabilities}) probabilities, and such visualisations appear, respectively, in Figs.~\ref{fig:vacuum_D-CTCs_CV_probabilities} and \ref{fig:vacuum_D-CTCs_CR_probabilities}. For both the CV and CR systems, these plots depict the transition from constant probability profiles ($\Vacuum=1$) to curved ones ($\Vacuum<1$), where the trough and peak behaviours become more pronounced as $\Vacuum\rightarrow0$. For the CV probabilities, an interesting feature is that while they are both exactly $\frac{1}{3}$ (which corresponds to the maximally entropic selection of $\Free$ for our qubit clocks) for $\Vacuum=1$, any infinitesimal variation of $\Vacuum$ below $1$ results in the probabilities jumping discontinuously to assume their curved behaviours. This indicates that the spectrum of clocks at differing rotation [as per the series in Eq.~(\ref{eq:vacuum_D-CTCs_CV})] materialises inside the CTC as soon as $\Vacuum$ varies from $1$.

\subsection{P-CTCs qubit visualisations}

\begin{figure}
	\includegraphics[scale=1.05]{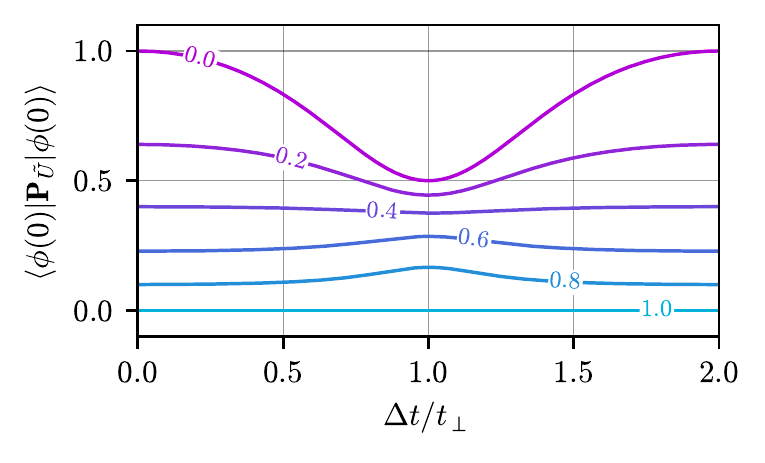}
	\includegraphics[scale=1.05]{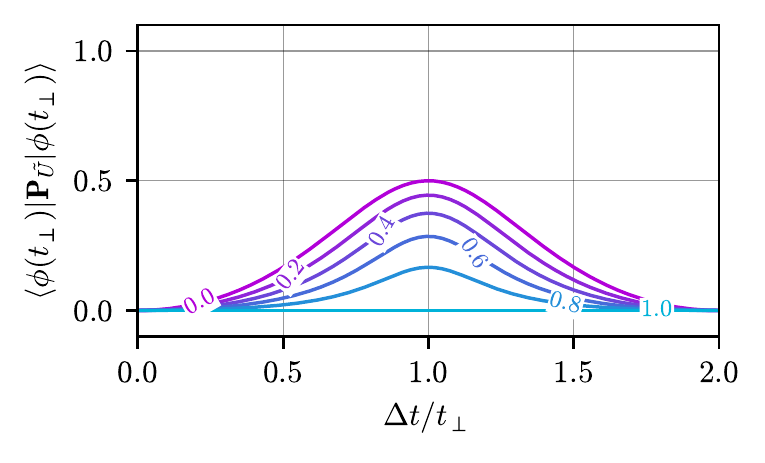}
	\vspace{-0.35cm}
	\caption{\label{fig:vacuum_P-CTCs_probabilities}P-CTCs clock probabilities (unitless) for the vacuum model plotted against the dimensionless time ratio. The lines correspond to varying values of $\sqrt{\Vacuum}$.}
\end{figure}

The qubit clock probabilities (\ref{eq:vacuum_P-CTCs_probabilities}) are visualised in Fig.~\ref{fig:vacuum_P-CTCs_probabilities}. Of interest is the fact that these probabilities behave in the same manner as the CR output from the D-CTCs analysis of Fig.~\ref{fig:vacuum_D-CTCs_CR_probabilities}, barring three relatively small differences. The first two of these is that the P-CTCs troughs (for the unevolved clock) and peaks (for the orthogonal clock) are both narrower and reach the same extrema (which occur at orthogonalisation). The third and perhaps most interesting difference is that the unevolved clock curves display a kind of behaviour inversion as $\Vacuum$ varies. Beginning initially with trough-like profiles at small $\Vacuum$, the lines transition into symmetric peaks (albeit small) at large $\Vacuum$ (e.g., $\sqrt{\Vacuum}=0.8$), until they finally flatten out at $\Vacuum=1$.

\begin{figure*}[t]
	\includegraphics[scale=1.05]{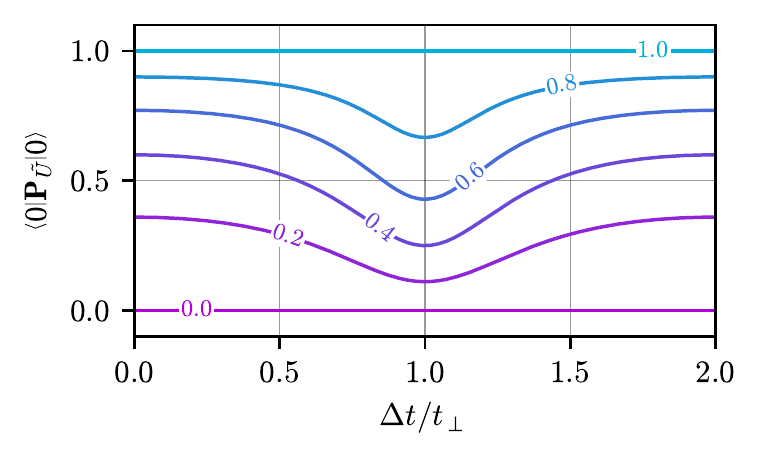}
	\includegraphics[scale=1.05]{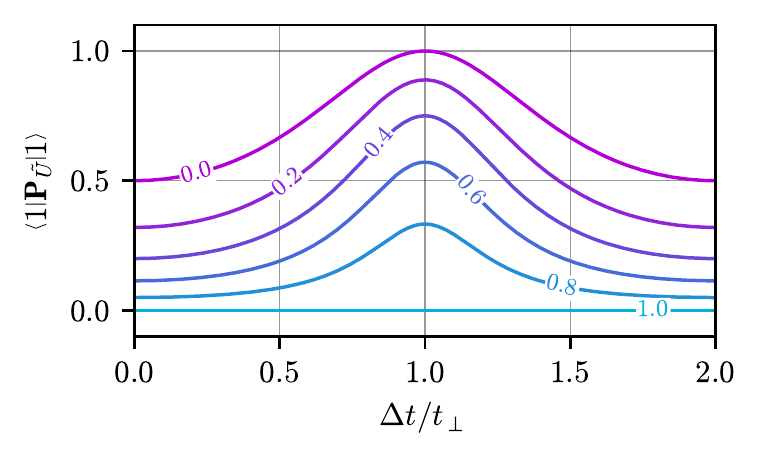}
	\includegraphics[scale=1.05]{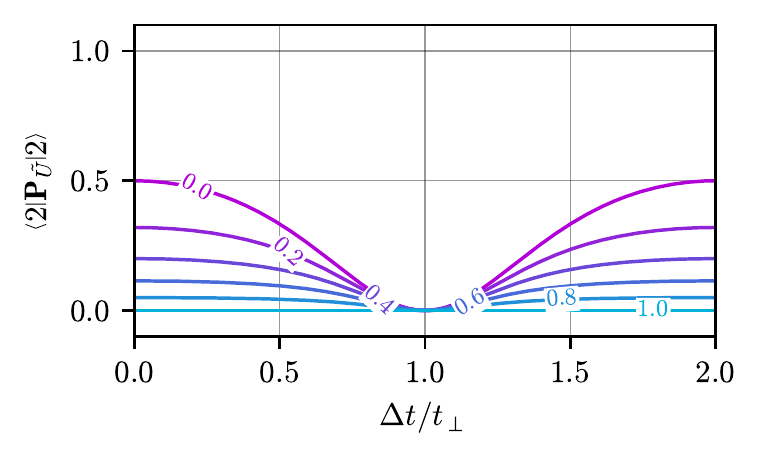}
	\vspace{-0.35cm}
	\caption{\label{fig:vacuum_P-CTCs_population}P-CTCs populations (unitless) for the vacuum model plotted against the dimensionless time ratio. The lines correspond to varying values of $\sqrt{\Vacuum}$.}
\end{figure*}

Next, the qubit plots appearing in Fig.~\ref{fig:vacuum_P-CTCs_population} serve to illustrate the behaviour of the P-CTCs populations (\ref{eq:vacuum_P-CTCs_populations}) in terms of $\Vacuum$ and $\DeltaTime$. Note the symmetry about the orthogonalisation time $\DeltaTime=\OrthogonalisationTime$ and that summation to $1$ is preserved for all $\Vacuum$ and $\DeltaTime$ (as expected). Also note how as $\Vacuum\rightarrow1$, the populations of the clock's nonvacuous energy states $\{\aket{n}\}_{n=1}^\NumberLevels$ go to zero, while the vacuum population goes to $1$. Additionally, the visualisations indicate how the clock's collective level amplitudes skew towards the ground state $\aket{1}$ as $\DeltaTime\rightarrow\OrthogonalisationTime$ from either side.

\hfill
\newpage

\bibliographystyle{apsrev4-2}
\interlinepenalty=10000
\bibliography{paper_1}

\end{document}